\documentclass[1p,times]{elsarticle}

\usepackage{ecrc}

\volume{00}

\firstpage{1}

\journalname{Physics Letters B}

\runauth{}

\jid{procs}

\jnltitlelogo{Physics Letters B}

\usepackage{amssymb}
 \usepackage{amsthm}

\usepackage[figuresright]{rotating}

\begin{document}

\begin{frontmatter}

\dochead{}

\title{Determining reaction cross sections via characteristic X-ray detection: $\alpha$-induced reactions on $^{169}$Tm for the astrophysical $\gamma$-process}

\author[label1]{G. G. Kiss \footnote{Present Address: Laboratori Nazionali del Sud, INFN, Catania, Italia}}

\ead{ggkiss@atomki.hu, Fax: (+36)(52) 416 181}

\author[label2]{T. Rauscher}

\author[label1]{T. Sz\"ucs}

\author[label1]{Zs. Kert\'esz}

\author[label1]{Zs.\,F\"ul\"op}

\author[label1]{Gy.\,Gy\"urky}

\author[label3]{C. Fr\"ohlich}

\author[label1]{J.\,Farkas}

\author[label1]{Z. Elekes}

\author[label1]{E.\,Somorjai}

\address[label1]{Institute of Nuclear Research (ATOMKI), H-4001 Debrecen, Hungary}
\address[label2]{Department of Physics, University of Basel, CH-4056 Basel, Switzerland}
\address[label3]{Department of Physics, North Carolina State University, Raleigh, NC 27695, USA}

\begin{abstract}
The cross sections of the $^{169}$Tm($\alpha$,$\gamma$)$^{173}$Lu and $^{169}$Tm($\alpha$,n)$^{172}$Lu reactions have been measured first time using a new method, by detecting the characteristic X-ray radiation following the electron capture-decay of $^{172,173}$Lu. Despite the relatively long half-life of the reaction products (T$_{1/2}$ = 500 and 6.7 days, respectively) it was possible to measure the cross section of the $^{169}$Tm($\alpha$,$\gamma$)$^{173}$Lu reaction close to the Gamow window (T$_9$ = 3.5 GK), between E$_\mathrm{c.m.}$ = 13.16 and 17.08 MeV. The $^{169}$Tm($\alpha$,n)$^{172}$Lu reaction cross section was measured from E$_\mathrm{c.m.}$ = 11.21 MeV up to E$_\mathrm{c.m.}$ = 17.08 MeV. The experimental results have been compared to theoretical predictions.
\end{abstract}


\end{frontmatter}

\section{Introduction}
\label{int}

In stellar nucleosynthesis involving charged particles two probability distributions are playing crucial role: the Coulomb penetration probability and the Maxwell--Boltzmann
energy distribution. The first strongly increases with increasing particle energy, but at a given temperature the probability of a particle having high energy falls off rapidly, following the Maxwell--Boltzmann distribution. The Coulomb penetration probability folded with the Maxwell--Boltzmann distribution forms the so-called Gamow peak, and the nuclear reactions are occurring in this narrow energy window \cite{ilibook,rau10}.
In order to provide reliable nuclear data for the astrophysical calculations, the experiments have to be performed in the Gamow window or at energies as close as possible. At higher burning temperatures this energy window can be shifted up to higher energies, however, it remains mostly well below the Coulomb barrier, making it hard or even impossible to perform experiments at astrophysically relevant energies \cite{ilibook}. Here -- since experimental data is missing in the most relevant mass range for the so-called astrophysical $\gamma$-process, too -- we propose a new experimental approach to measure the cross sections of important nuclear reactions.

\begingroup
\begin{table}
\center
\caption{\label{tab:activation} Existing experimental ($\alpha$,$\gamma$) database relevant for the astrophysical $\gamma$-process measured via activation method and $\gamma$-counting in the A $\geq$ 100 region. The target nuclei, the half-lives of the reaction products, the Gamow windows \cite{rau10} and the lowest energies measured (E$_\mathrm{min}$) are given.
}
\begin{tabular}{cccccc}
\multicolumn{1}{c}{Target} &
\multicolumn{1}{c}{Half-} &
\multicolumn{1}{c}{Gamow window} &
\multicolumn{1}{c}{E$_{min}$} &
\multicolumn{1}{c}{Ref.} \\
\multicolumn{1}{c}{nucleus} &
\multicolumn{1}{c}{life [d]} &
\multicolumn{1}{c}{[MeV] (T$_9$ = 3.5 GK)} &
\multicolumn{1}{c}{[MeV]} &
\multicolumn{1}{c}{} \\
\hline
 $^{106}$Cd             & 0.17 & 6.6 - 10.1 & 7.57   & \cite{gyu06} \\
 $^{112}$Sn             & 0.10 & 6.8 - 10.7 & 9.97   & \cite{ozk07,rap08} \\
 $^{113}$In             & 0.12 & 6.5 - 9.4 & 8.66   & \cite{yal09} \\
 $^{117}$Sn             & 154  & 6.1 - 9.6 & 11.9   & \cite{cat08} \\
 $^{197}$Au             & 3.05 & 8.5 - 11.7& 17.5  & \cite{bas07} \\
 $^{169}$Tm             & 500.5& 8.2 - 11.3 & 13.2   & this work \\
\end{tabular}
\end{table}
\endgroup

In order to understand the synthesis of the so-called $p$ nuclei ---  the rare proton rich isotopes between   $^{74}$Se and $^{196}$Hg \cite{woo78} --- experimental data at sub-Coulomb energies  are particularly important. Despite the tremendous experimental and theoretical efforts of recent years, the synthesis of the $p$ nuclei is still one of the least known processes of nucleosynthesis. It is generally accepted, however, that the main stellar mechanism synthesizing these nuclei --- the so called $\gamma$-process --- 
is initiated by ($\gamma$,n) photodisintegration reactions on preexisting more neutron-rich seed nuclei. As the neutron separation energy increases along the ($\gamma$,n) path towards more neutron deficient isotopes, ($\gamma$,p) or ($\gamma$,$\alpha$) reactions become faster and process the material towards lower masses \cite{arn03,rau06,rap06}. Recently, consistent studies of the nucleosynthesis of the $p$ nuclei become available, employing theoretical reaction rates in large reaction networks \cite{rau06,rap06}, proving that in the case of the production of the heavy $p$ nuclei (140 $\leq A \leq$ 200) the 
reaction flow is strongly sensitive to the ($\gamma$,$\alpha$) photodisintegration rates. If those reaction rates are high, more material will contribute to the synthesis of lower mass $p$ isotopes. On the other hand, if the reaction rates are lower, processing toward lower atomic numbers is weaker, resulting in a relative enrichment in the higher mass $p$ isotopes such as $^{174}$Hf, $^{180}$W, and $^{190}$Pt \cite{rap06}. Consequently, to reproduce the path of the $\gamma$-process, experimental data is highly needed in this mass range.

\section{Experimental approaches used in $\gamma$-process studies}
\label{app}

In principle, photodisintegration cross sections can be determined directly by photon induced reaction studies \cite{nai08}. However, in such an experiment the target nucleus is always in its ground state, whereas in stellar environments thermally populated excited states also contribute to the reaction rate leading to large corrections of the ground state rate which can only be modeled theoretically \cite{moh08}. It has been shown that the influence of thermal population is much less pronounced in capture reactions and therefore it is advantageous to measure in the direction of capture and convert the measured rate to the rate of the inverse reaction by applying the principle of detailed balance \cite{supplett,suppprc}. Recently, several ($\alpha$,$\gamma$) cross sections around A $\approx$ 100 have been measured using the activation method \cite{ful96,rap02,gyu06,ozk07, rap08,yal09,cat08} (see Table \ref{tab:activation}). Above the A $\approx$ 100 mass region, however, there are practically no experimental ($\alpha$,$\gamma$) data below the Coulomb barrier except the $^{144}$Sm($\alpha$,$\gamma$)$^{148}$Gd reaction \cite{som98}. The results of this latter experiment (using a method being limited by the fact that the yield of emitted $\alpha$-particles had to be measured) showed that there are large discrepancies between the experimental results and theoretical predictions \cite{som98}.

Though the activation method proves to be successful to measure $\alpha$-induced cross sections it has numerous limitations which can be sorted into two groups (physical and technical limitations):

(1) In order to have reasonable count rates, reactions producing nuclei with long half-life (more than few days) were typically excluded. Furthermore, the product nuclei must decay via the emission of at least one high intensity (typically 70-97\%) $\gamma$ transition.

(2) A detector with a good signal-to-noise ratio is necessary for going to low energies. A high efficiency detector must be used to compensate the low cross sections, long half-lives or unfavorable $\gamma$-branching ratios. The laboratory background can be reduced using lead shielding but the beam-induced background on the impurities of the target and/or the backing will still be present. Furthermore, if the reaction product decays via $\gamma$ cascades, the true coincidence summing effect has to be taken into account.

\begingroup
\begin{table}
\center
\caption{\label{tab:decay}Decay parameters of the $^{169}$Tm($\alpha$,$\gamma$)$^{173}$Lu and $^{169}$Tm($\alpha$,n)$^{172}$Lu reaction products \cite{nndc}.}
\begin{tabular}{ccccc}
\multicolumn{1}{c}{Residual} &
\multicolumn{1}{c}{Decay} &
\multicolumn{1}{c}{Half-} &
\multicolumn{1}{c}{Energy} &
\multicolumn{1}{c}{Relative intensity} \\
\multicolumn{1}{c}{nucleus} &
\multicolumn{1}{c}{mode} &
\multicolumn{1}{c}{life [d]} &
\multicolumn{1}{c}{[keV]} &
\multicolumn{1}{c}{per decay [\%]} \\
\hline
$^{173}$Lu  & $\epsilon$ 100\% & 500  $\pm$ 3.65 & 51.35 (K$_{\alpha_2}$)   & 43.8 $\pm$ 1.4 \\
            &                  &                 & 52.39 (K$_{\alpha_1}$)   & 76.3 $\pm$ 2.4 \\
$^{172}$Lu  & $\epsilon$ 100\% &  6.7 $\pm$ 0.04 & 51.35  (K$_{\alpha_2}$)  & 31.5 $\pm$ 0.9 \\
            &                  &                 & 52.39  (K$_{\alpha_1}$)  & 54.9 $\pm$ 1.5 \\
            &                  &                 & 810.06                & 16.6 $\pm$ 0.9 \\
            &                  &                 & 900.72                & 29.8 $\pm$ 1.3 \\
            &                  &                 & 912.08                & 15.3 $\pm$ 0.7 \\
            &                  &                 & 1093.63               & 63.0 $\pm$ 3.0 \\
\end{tabular}
\end{table}
\endgroup

In this work the cross sections of the $^{169}$Tm($\alpha$,$\gamma$)$^{173}$Lu and $^{169}$Tm($\alpha$,n)$^{172}$Lu reactions have been measured, by detecting the characteristic K$_{\alpha1-2}$ X-ray lines following the electron capture-decay of produced Lutetium isotopes. It has to be realized that unstable nuclei above the mass A $\approx$ 150 on the proton-rich side close to the valley of stability typically decay by electron capture. Such an electron capture is usually followed by X-ray emission. The characteristics X-ray counting approach has numerous advantages: the relative intensities are high since the electron capture is followed dominantly by the emission of either a K$_{\alpha1}$ or a K$_{\alpha2}$ X-ray (the probability of Auger electron or K$_{\beta}$ X-ray emission are typically an order of magnitude smaller). Furthermore, after the decay of one product nucleus only one K$_{\alpha}$ characteristic X-ray is emitted and consequently the target can be put close to the surface of the detector the summing effect is not present. 

To measure these X-ray lines a so-called LEPS  (Low Energy Photon Spectrometer) detector -- consisting of a thin germanium crystal with large surface and a thin entrance window -- was used. This detector has high efficiency for the K$_{\alpha}$ lines and insensitive to the high energy $\gamma$-s of the background. Consequently -- compared to a standard shielded HPGe detector -- the observable minimum counting rates were orders of magnitude smaller (see Fig. 1).  This way, despite the relatively long half-life of the reaction product (T$_{1/2}$ = 500 days) it was possible to measure the cross section of the $^{169}$Tm($\alpha$,$\gamma$)$^{173}$Lu reaction close to the astrophysically important energy region. The $^{169}$Tm($\alpha$,n)$^{172}$Lu (T$_{1/2}$ = 6.7 days)  reaction was studied between  E$_\mathrm{c.m.}$ = 11.21 MeV up to E$_\mathrm{c.m.}$ = 17.08 MeV.

The disadvantage of the X-ray counting is that this approach is not isotope selective, e.g. in this case it is impossible to distinguish between the decays of the different lutetium isotopes produced via the $^{169}$Tm($\alpha$,$\gamma$)$^{173}$Lu and the $^{169}$Tm($\alpha$,n)$^{172}$Lu reactions. However,  by taking the advantage of the vastly different half-lives it is possible to overcome this problem. Namely, the X-ray counting was carried out at least two times: at first after the end of the irradiation to determine the  $^{169}$Tm($\alpha$,n) cross section and about 24 weeks later when the $^{172}$Lu activity of the targets drastically reduced and the measured X-ray yield belongs solely to the decay of the $^{173}$Lu, produced via the $^{169}$Tm($\alpha$,$\gamma$) reaction (for details, see below). 

\section{Experimental technique}
\label{exp}

\begin{figure}
\center
\resizebox{0.8\columnwidth}{!}{\rotatebox{0}{\includegraphics[clip=]{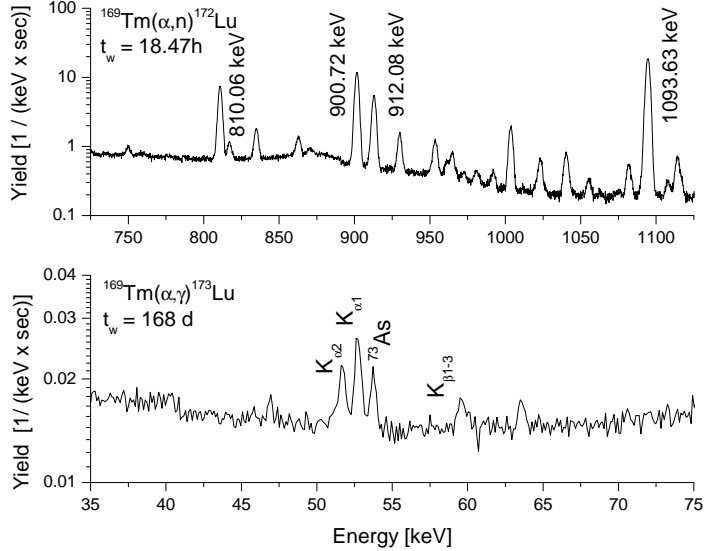}}}
\caption{\label{fig:spectra} Off-line $\gamma$- (upper panel) and characteristic X-ray (lower panel) spectra normalized to the length of the countings, taken after irradiating a $^{169}$Tm target with 15.5 MeV $\alpha$ beam. The $\gamma$-lines used to measure the cross section of the $^{169}$Tm($\alpha$,n) reaction and the characteristic X-ray radiation used to determine the $^{169}$Tm($\alpha,\gamma$) reaction cross section are marked.}
\end{figure}

Here we provide a proof of concept by measuring the ($\alpha$,$\gamma$) and ($\alpha$,n) reactions on $^{169}$Tm at low energies. The targets have been made by evaporating metallic thulium onto 2 $\mu$m thick Al foils. The thickness of the thulium targets varied between 106 $\mu$g/cm$^2$ and 356 $\mu$g/cm$^2$, at higher energy irradiations thinner and at lower energy irradiations thicker targets were used.
The absolute number of target atoms and the uniformity were determined by the PIXE method (Proton Induced X-ray Emission) \cite{pixe} using the Nuclear Microbeam Facility of ATOMKI. A 2 MeV proton beam -- the beamspot was 3 $\mu$m x 3 $\mu$m -- was scanned over a surface of 500 $\mu$m x 500 $\mu$m at several different positions on the target. The precision of the determination of the number of target atoms was better than 3\% and the thickness was found to be uniform within 3\%. Weighing was also used to determine the number of target atoms. The agreement between the results of the two methods is within 4\%. The $^{169}$Tm targets have been irradiated with $\alpha$ beams from the MGC cyclotron of ATOMKI. The energy of the $\alpha$ beam was between E = 11.5 MeV - 17.5 MeV, covered with 0.5 MeV - 1 MeV steps with beam currents of typically 2 $\mu$A. After the beam-defining aperture, the chamber was insulated and secondary electron suppression voltage of $-300$ V was applied at the entrance of the chamber. The collected charge in each irradiation was between 50 mC and 250 mC. After irradiation, T$_{waiting}$ = 1h - 3h  waiting time was used in order to let the short-lived, disturbing activities decay.

\begingroup
\begin{table}
\center
\caption{\label{tab:results} Experimental $S$ factors of the $^{169}$Tm($\alpha$,n)$^{172}$Lu and $^{169}$Tm($\alpha$,$\gamma$)$^{173}$Lu reactions.}
\begin{tabular}{cccc}
\multicolumn{1}{c}{E$_\mathrm{lab}$} &
\multicolumn{1}{c}{E$_\mathrm{c.m.}$} &
\multicolumn{1}{c}{$^{169}$Tm($\alpha$,n)$^{172}$Lu} &
\multicolumn{1}{c}{$^{169}$Tm($\alpha$,$\gamma$)$^{173}$Lu}\\
\multicolumn{1}{c}{$\left[ MeV\right]$} &
\multicolumn{1}{c}{$\left[ MeV\right]$} &
\multicolumn{1}{c}{$\left[ 10^{28} MeV barn \right]$} &
\multicolumn{1}{c}{$\left[ 10^{25}MeV barn \right]$} \\
\hline
11.5  & 11.21 $\pm$ 0.057 &  65.7 $\pm$ 6.9   &   \\
11.85 & 11.55 $\pm$ 0.059 & 53.5 $\pm$ 5.8   &   \\
12.2  & 11.90 $\pm$ 0.061 &  46.5 $\pm$ 5.3   &   \\
12.5  & 12.19 $\pm$ 0.062 &  36.6 $\pm$ 2.6   &   \\
13.5  & 13.16 $\pm$ 0.066 &  24.0 $\pm$ 1.7   &  329     $\pm$ 33  \\
14.0  & 13.66 $\pm$ 0.069 &  18.6 $\pm$ 1.4   &  184     $\pm$ 16  \\
15.0  & 14.63 $\pm$ 0.075 &  12.0 $\pm$ 0.9   &  52.9    $\pm$ 5.2    \\
15.5  & 15.12 $\pm$ 0.077 &  9.48 $\pm$ 1.01   &  31.0  $\pm$ 3.2  \\
16.0  & 15.61 $\pm$ 0.079 &  6.04 $\pm$ 0.49   & 16.1   $\pm$ 1.2   \\
16.5  & 16.10 $\pm$ 0.081 &  4.51 $\pm$ 0.48   & 7.87   $\pm$ 0.61\\
17.0  & 16.59 $\pm$ 0.084 &  3.27 $\pm$ 0.29   & 4.11   $\pm$ 0.33\\
17.5  & 17.08 $\pm$ 0.086 &  2.23 $\pm$ 0.20   & 2.40   $\pm$ 0.18\\
\end{tabular}
\end{table}
\endgroup

The decay parameters of $^{172,173}$Lu taken from literature are summarized in Table \ref{tab:decay}. The activity of all samples has been measured with a LEPS  detector (GL2015R type, produced by Canberra). The characteristic X-ray countings were carried out using two geometries: after the irradiation, the targets were put 12 cm away from the detector (far geometry). The countings were repeated several times, 1 - 4 weeks later in close geometry, in these cases the target detector distance was 5 cm. The absolute efficiency of the LEPS detector was determined using the 53.16 keV line of $^{133}$Ba and the 59.54 keV line of the $^{241}$Am calibrated radioactive sources in far geometry where the true coincidence effect is negligible. A linear fit for the measured absolute efficiencies were used to determine the absolute efficiencies in the 53 keV - 60 keV energy range. Taking into account the elapsed time and the decay between the countings, a conversion factor between the two geometries was calculated and used to derive the absolute efficiency in close geometry.

At higher bombarding energies (at and above 15 MeV) the $\gamma$-activity of the targets has also been measured with a 40\% relative efficiency HPGe detector, while at 13.5 MeV and 14 MeV a 100\% relative efficiency HPGe detector in Ultra Low Background (ULB) configuration has been used. The cross section of the ($\alpha$,n) reaction obtained from the measurements with the LEPS and the HPGe detectors were found to be in agreement within 4\%. $\gamma$-lines belonging to the $^{169}$Tm($\alpha,\gamma$)$^{173}$Lu reaction were not observable in the $\gamma$-spectra taken either with the ULB or the HPGe detectors. 
\footnote{An attempt was also made to measure the low activity of some targets in an ultra low background environment in the LNGS underground laboratory in Italy. The analysis of these measurements is still in progress and the results will be published elsewhere. }

The cross sections of the $^{169}$Tm($\alpha$,$\gamma$)$^{173}$Lu reaction were determined by counting the yield of the K$_{\alpha_1}$ and K$_{\alpha_2}$ characteristic X-ray lines using the LEPS in close geometry.  Figure \ref{fig:spectra} shows typical spectra measured with the HPGe detector (upper panel, t$_\mathrm{waiting}$ = 18.47 h) and the LEPS (lower panel, t$_\mathrm{waiting}$ = 168 d) after irradiating the $^{169}$Tm target with a 15.5 MeV $\alpha$ beam. In the measured X-ray spectra it is impossible to distinguish between events coming from the decay of the different Lu isotopes. However, there is a large difference in their half-lives and therefore we can assume that -- even if the cross sections were the same -- the yield measured after the irradiation would be dominated by the decay of $^{172}$Lu. To determine the cross section of the $^{169}$Tm($\alpha$,$\gamma$) reaction, the activity measurement using the LEPS was repeated twice, $\approx$ 168 days and 200 days after irradiating the targets with $\alpha$ beam. During the cooling period of more than 24 weeks the $^{172}$Lu activity of the targets decreased by a factor of more than 3 x 10$^7$, therefore the observed X-ray yield belongs solely to the decay of the $^{173}$Lu produced only by the $^{169}$Tm($\alpha$,$\gamma$) reaction. The ($\alpha$,$\gamma$) cross sections obtained in the 2 measurements are within 3\%. 
The target material contained Fe and Ge impurities. The 73 keV and 122 keV background peaks, originating from the decay of the reaction products of $^{70}$Ge($\alpha$,n)$^{73}$Se and $^{54}$Fe($\alpha$,n)$^{57}$Ni, were used to monitor the detector efficiency. No long term variation in the efficiency was found.

\begin{figure}
\center
\resizebox{0.8\columnwidth}{!}{\rotatebox{0}{\includegraphics[clip=]{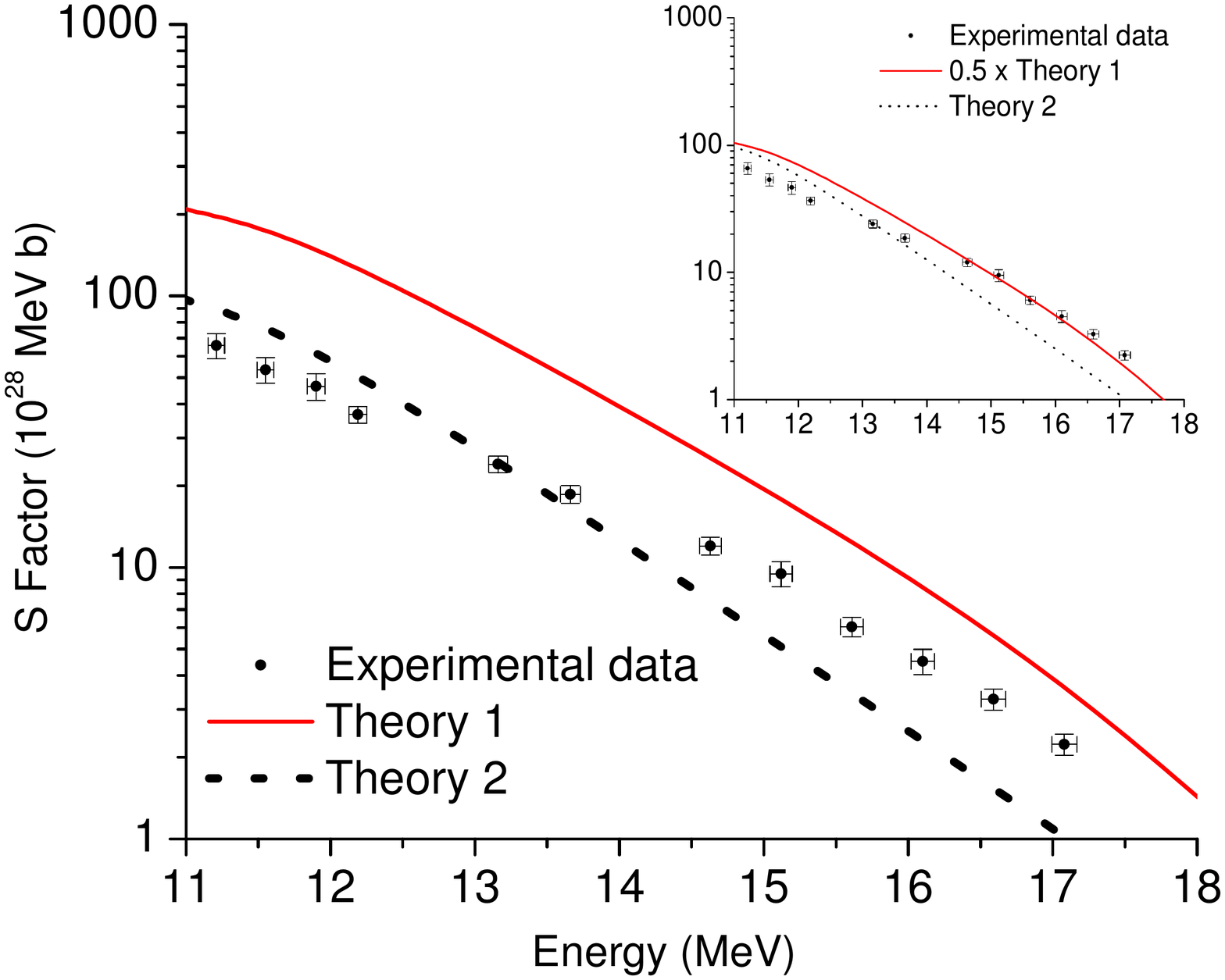}}}
\caption{\label{fig:an} Experimental $^{169}$Tm($\alpha$,n) $S$ factors (exp) are compared to two theoretical calculations. Both calculations use the same code \cite{nonsmokerweb} with the same settings except for the optical $\alpha$+nucleus potential. One calculation was performed
with the standard potential by \cite{mcf66} (Theory1), indicated by solid line (the inset shows the result divided by factor of 2, see text). Alternatively, the recent Fr\"ohlich potential \cite{frodip,fro02} was used (Theory2). Plotted is the \textit{original, unrenormalized} result (dashed line).}
\end{figure}

\section{Results and discussion}
\label{dis}

In astrophysical investigations it is common to quote the astrophysical $S$ factor. The cross section, $\sigma$(E) and the astrophysical $S$ factor, $S$(E) at c.m. energy $E$ are related by:

\begin{equation}
 S(E)=\frac{E\sigma(E)} {exp(2\pi\eta)}, 
\end{equation}

with $\eta$ being the Sommerfeld parameter. Several irradiations at the same energies were repeated, the astrophysical $S$ factors shown in Table \ref{tab:results} were derived from the averaged results of the irradiations weighted by the statistical uncertainty of the measured values. The effective center of mass energies (E$_\mathrm{c.m.}$) in the second
column take into account the energy loss of the $\alpha$ beam in the target. The quoted uncertainty in the E$_\mathrm{c.m.}$ values corresponds to the energy stability of the $\alpha$ beam and to the uncertainty of the energy loss in the target. The uncertainty of the $S$ factors is the quadratic sum of the following partial errors: efficiency of the HPGe detector and LEPS (6\% and 4\%, respectively), number of target atoms (4\%), current measurement (3\%), uncertainty of decay parameters ($\leq$\ 5 \%) and counting statistics (0.5 - 7\%).

\begin{figure}
\center
\resizebox{0.8\columnwidth}{!}{\rotatebox{0}{\includegraphics[clip=]{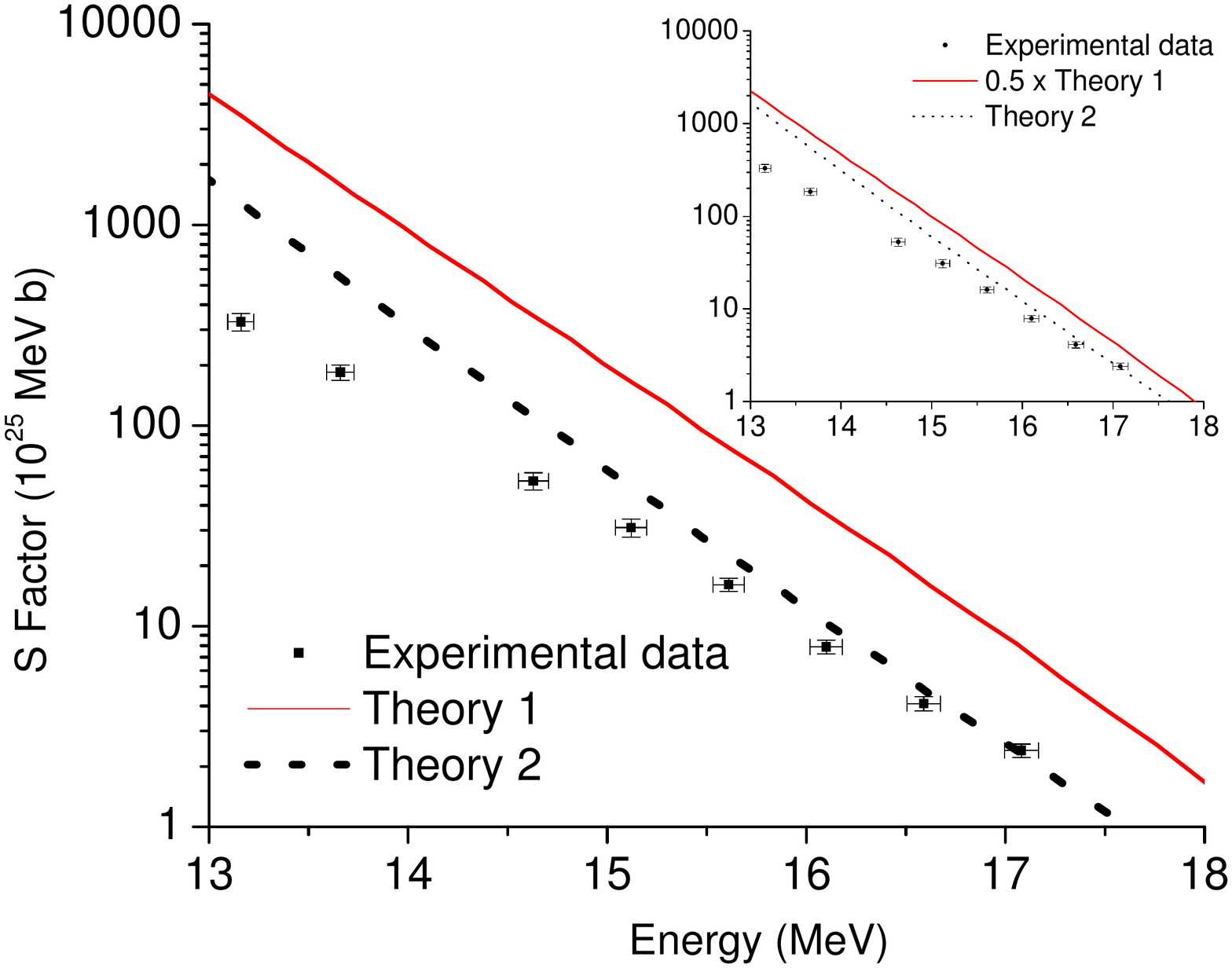}}}
\caption{\label{fig:ag} Same as Fig.\ \ref{fig:an} for $^{169}$Tm($\alpha$,$\gamma$)$^{173}$Lu.}
\end{figure}

The experimental $S$ factors have been compared to predictions of the statistical model (Hauser--Feshbach) code NON-SMOKER$^\mathrm{WEB}$ v5.8.1 \cite{nonsmokerweb}. This code is a further development of the NON-SMOKER statistical model code \cite{NONSMOKER} and
includes, among other modifications, updated masses and excited states of nuclei.
The results underline the importance of measurements at energies as close as possible to the astrophysically relevant energy window. The reaction
cross sections are sensitive to different nuclear properties depending on the projectile energy and reaction type. It is already known that the standard
optical $\alpha$+nucleus potential by \cite{mcf66}, widely used in astrophysical calculations, may not be optimal at $\alpha$ energies close to or
below the Coulomb barrier. This can be seen in Fig.\ \ref{fig:an} showing $S$ factors for the ($\alpha$,n) channel. In the measured energy range, the
theoretical $S$ factors for this reaction are mainly sensitive to the calculated average $\alpha$ widths which, in turn, mainly depend on the optical
potential used. The calculation using the standard potential (Theory1) overestimates the cross section by a factor of 2 (this is demonstrated on the inset of Fig.\ \ref{fig:an}).
On the other hand, the calculation using the potential of \cite{frodip,fro02} (Theory2) is much closer to our experimental values, especially at low energies, without any renormalization. This confirms earlier findings that this potential describes low-energy $\alpha$-induced reactions better although it does not reproduce scattering data. This puzzle needs
further investigation both theoretically and experimentally.

The $^{169}$Tm($\alpha$,$\gamma$) $S$ factors (Fig.\ \ref{fig:ag}) are not described well with either potential. However, the values obtained with the
potential of \cite{frodip,fro02} (Theory2) are again much closer to the experimental values than the ones obtained with the standard potential \cite{mcf66}
(Theory1). Contrary to the ($\alpha$,n) reaction, the ($\alpha$,$\gamma$) reaction cross sections are not only sensitive to the optical potential but
also to the calculated $\gamma$ width. Since the ($\alpha$,n) $S$ factors are described well with the potential of \cite{frodip,fro02}, we can assume
that the remaining deficiencies in the reproduction of the ($\alpha$,$\gamma$) excitation function are due to the calculated average $\gamma$ width. This width
itself depends on the GDR properties and the level density of $^{173}$Lu, which are both experimentally undetermined. However, the ($\alpha$,$\gamma$) reaction rate at $p$ process temperatures is insensitive to the $\gamma$ width and therefore an inadequate modeling is astrophysically inconsequential.

\section{Summary}
\label{sum}

In the present work the cross section of the $^{169}$Tm($\alpha$,$\gamma$) and $^{169}$Tm($\alpha$,n) reactions were determined well below the Coulomb barrier with high precision ($\leq$ 10\% total uncertainty) in spite of the very long half-life (500 and 6.7 d) of the reaction products by counting the characteristic X-ray radiation. It has to be emphasized that this method is less sensitive to the usual obstacles of the activation technique such as low $\gamma$ branching ratios or long half-lives. Moreover, not only ($\alpha,\gamma$) reactions in the region of the heavy $p$ nuclei can be studied via this X-ray radiation approach but it can be used also for other applications provided the reaction products decay via electron-capture, e.g., in ($\gamma$,n), ($\gamma$,p) and ($\gamma$,$\alpha$) photoactivation studies.

\section{Acknowledgements}

We thank the valuable comments and suggestions of B. Sulik and G. Kalinka. 
This work was supported by the EUROGENESIS research
program, the European Research Council grant agreement no. 203175, the Economic Competitiveness Operative Programme GVOP-3.2.1.-2004-04-0402/3.0., OTKA (NN83261, K68801),
and the Swiss NSF (grant 2000-105328). 

\bibliographystyle{elsarticle-num}

\end{document}